\documentclass{article}
\usepackage{spconf,amsmath,graphicx}
\usepackage[hidelinks]{hyperref}
\ninept
\usepackage{textcomp}
\usepackage{subcaption}
\usepackage{url}
\usepackage{algorithmic}
\usepackage{cite}
\usepackage{amsmath,amssymb,amsfonts}
\usepackage{placeins}

\makeatletter
\def\thebibliography#1{\section{References}\list
 {[\arabic{enumi}]}{\settowidth\labelwidth{[#1]}\leftmargin\labelwidth
 \advance\leftmargin\labelsep
 \itemsep=0pt\relax\parsep=0pt\relax\topsep=2pt\relax
 \usecounter{enumi}}
 \def\newblock{\hskip .11em plus .33em minus .07em}
 \sloppy\clubpenalty4000\widowpenalty4000
 \sfcode`\.=1000\relax}

\makeatother

\title{Interpreting Content and Speaker Characteristics in Factorised Self-Supervised Subspaces}
\name{Kyle Janse van Rensburg$^1$, Benjamin van Niekerk$^2$, Herman Kamper$^1$\thanks{Work supported by the National Research Foundation (NRF) of South Africa and the Department of Science, Technology and Innovation (DSTI).}}
\address{$^1$E\&E Engineering, Stellenbosch University \quad $^2$Concordia University and Mila Quebec AI Institute}
\usepackage{xcolor}
\definecolor{mycolor}{HTML}{FF6600}
\definecolor{coolcolor}{HTML}{26209e}
\definecolor{plancolor}{HTML}{aaa6ed}
\definecolor{nelcolor}{HTML}{3060F2}

\begin{document}

\maketitle
\begin{abstract}
Self-supervised speech features encode both content and speaker information.
Recent work introduced an SVD-based factorisation method that decomposes these features into a shared content matrix capturing temporal variation and speaker-specific transformations capturing static speaker characteristics.
However, how information is organised within these components remains unclear.
In this paper, we investigate how the dimensions of WavLM-factorised content and speaker subspaces correlate with speech characteristics such as pitch, intensity, and voicing.
We find that leading dimensions in the content space primarily capture intensity, higher-order formants, and voicing, while pitch is encoded in a later dimension.
In contrast, the highest-variance speaker dimension is strongly associated with pitch and gender, with later dimensions capturing high-frequency variation.
Intervention experiments show that manipulating these dimensions enables targeted control of speech characteristics for speech synthesis.
Furthermore, modifying the content and speaker representations jointly provides fine-grained control over characteristics such as pitch and intensity.
\end{abstract}
\begin{keywords}
interpretability, disentanglement, self-supervised learning, speech representations, voice manipulation
\end{keywords}
\section{Introduction}
\label{sec:Intro}

Representations obtained through self-supervised learning (SSL) have become standard in speech processing.
Thorough analyses of these features~\cite{pasad2021,pasad2023} show that both content and speaker information are encoded, enabling downstream tasks from speech recognition~\cite{chiu2022} to speaker verification~\cite{lepage2025} to voice conversion~\cite{kNN-VC}.
This is despite the entanglement of speaker and content in SSL representations.

Recently, an approach has been proposed to linearly factorise SSL features into content and speaker subspaces~\cite{LinearVC}.
Using singular value decomposition (SVD), the method identifies a shared, low-dimensional content matrix capturing temporal variation common across speakers, alongside speaker-specific linear transformations encoding static speaker identity.
The method requires no training beyond solving a least-squares problem.

Although successfully applied to voice conversion~\cite{LinearVC} and subsequently extended~\cite{USCF}, the internal structure of the resulting subspaces remains poorly understood: it is unclear how information is organised within the content and speaker matrices, or whether individual dimensions carry meaningful, interpretable information.
Addressing this is important both for understanding what SSL representations encode and for enabling controllable speech processing.

To answer these questions, we extract SSL features from WavLM~\cite{WavLM} and factorise them into content and speaker subspaces.
The content matrix can be analysed directly, as its columns are ordered by singular value.
The speaker matrices are not so simple, so we flatten them and apply principal component analysis (PCA) to identify the directions of greatest variation across speakers.
We then correlate specific speech characteristics (e.g., pitch, gender, intensity, voicing) with the dimensions of each subspace.

Our correlation analysis shows that leading content matrix dimensions primarily capture intensity, higher-order formants, and voicing, with pitch appearing only in a later dimension.
The highest-variance speaker space dimension, however, is strongly associated with pitch and gender, with later dimensions capturing high-frequency variation.
Broadly speaking, temporally-varying characteristics are captured by the content space and static characteristics by the speaker space, with characteristics tied to both (e.g., spectral rolloff, ZCR) correlating with dimensions in both subspaces.
Intervention experiments, in which we modify a dimension and resynthesise the vocoded audio, confirm that these dimensions give targeted, and jointly even finer-grained, control over speech characteristics such as pitch and intensity.
This opens up the possibility of simple, training-free voice control in future work.

\begin{figure*}[!t]
\centering
    \begin{subfigure}[b]{0.49\textwidth}
        \centering
        \includegraphics[width=\linewidth]{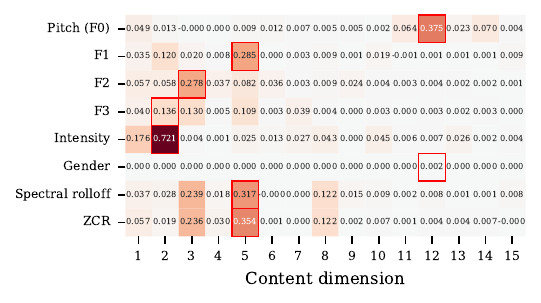}
        \caption{$\mathbf{C}$-space correlations between content dimensions and temporal speech characteristics.}
        \label{fig:C_correlations}
    \end{subfigure}
    \hfill
    \begin{subfigure}[b]{0.49\textwidth}
        \centering
        \includegraphics[width=\linewidth]{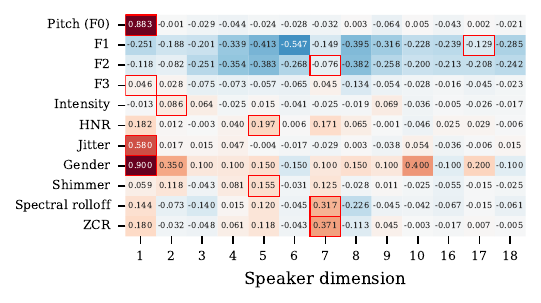}
        \caption{$\mathbf{S}$-space correlations between speaker dimensions and global speaker characteristics.}
        \label{fig:S_correlations}
    \end{subfigure}
\caption{
Heat maps showing correlation scores between speaker-specific characteristics and the dimensions of the factorised subspaces, using WavLM-Large layer six on development data.
}
\label{fig:correlation}
\end{figure*}

\section{Content Factorisation}
\label{sec:Content_fact}

We briefly review the content factorisation approach of~\cite{LinearVC}, which separates temporal content from global speaker information in SSL representations.
We also review the extension of~\cite{USCF}, needed to apply the approach to unseen speech and speakers.

\subsection{Content factorisation on a training set}
\label{subsec:content_fact_original}

Content factorisation splits an SSL representation into a content ($\mathbf{C}$) and a speaker ($\mathbf{S}$) subspace, such that~\cite{LinearVC}:
\begin{equation}
    \mathbf{X} \approx \mathbf{C}\mathbf{S}
\label{eq:content_fact}
\end{equation}
where $\mathbf{X} \in \mathbb{R}^{N \times D}$ are $D$-dimensional SSL frames, $\mathbf{C} \in \mathbb{R}^{N \times r}$ is a shared content matrix, and $\mathbf{S} \in \mathbb{R}^{r \times D}$ is a speaker-specific transformation.
Since $\mathbf{C}$ should represent content independent of who is speaking, it must be estimated from parallel speech (i.e., different speakers saying the same thing).
Lacking parallel data, \cite{LinearVC} instead selects $K$ training speakers and, from them, a random pivot speaker.
They then align $N$ pivot frames to every other speaker via nearest-neighbour lookup to obtain content-aligned frames $\mathbf{X}_k$ for speaker $k$.
Finally, they solve for a shared $\mathbf{C}$ and per-speaker matrices $\{\mathbf{S}_k\}$ via SVD by minimising $\sum_{k=1}^{K}||\mathbf{X}_k - \mathbf{C}\mathbf{S}_k||_F^2$ subject to $\text{rank}(\mathbf{C}\mathbf{S}_k)\leq r$.
We use $N=8192$ pivot frames and rank $r=64$.
This gives $\mathbf{C}$ and $\{\mathbf{S}_k\}$ only for the $K$ training speakers.
Therefore, we need two further steps to handle unseen speakers and unseen utterances.

\subsection{Finding $\mathbf{S}$ for unseen speakers}
\label{subsec:content_fact_artificial_S}

For a new speaker, we align their speech to the pivot frames as above to obtain $\mathbf{X}_{\text{new}}$, and solve for their speaker matrix by treating it as an unknown in a linear regression against the already-known $\mathbf{C}$: $\mathbf{S}_{\text{new}} \approx \mathbf{C}^{+}\mathbf{X}_{\text{new}}$, where $\mathbf{C}^+$ is the pseudo-inverse of $\mathbf{C}$.

\subsection{Finding $\mathbf{C}$ for unseen utterances}
\label{subsec:content_fact_per_utt_C}

To analyse an utterance at the frame level, we also need $\mathbf{C}$ for unseen speech from an unseen speaker.
Taking inspiration from~\cite{USCF}, we learn a transformation $\mathbf{W}$ and bias $\mathbf{b}$ that map any utterance to its content matrix:
\begin{equation}
    \mathbf{C}_{\text{inf}} \approx \mathbf{X}_{\text{inf}} \mathbf{W} + \mathbf{b}
\label{eq:per_utterance_W_and_b}
\end{equation}
where $\mathbf{X}_{\text{inf}} \in \mathbb{R}^{N \times D}$, $\mathbf{W} \in \mathbb{R}^{D \times r}$, and $\mathbf{b} \in \mathbb{R}^{1 \times r}$ (added row-wise).
$\mathbf{W}$ and $\mathbf{b}$ are fit by least squares on the training data, minimising $\sum_{k=1}^{K}||\mathbf{X}_k\mathbf{W} + \mathbf{b} - \mathbf{C}_{\text{align}}||_F^2$, where $\{\mathbf{X}_k\}$ are the pivot-aligned training frames and $\mathbf{C}_{\text{align}}$ is the known $\mathbf{C}$ from the original factorisation.
The fitted $\mathbf{W}, \mathbf{b}$ then apply directly to unseen utterances via~\eqref{eq:per_utterance_W_and_b}.

\section{Methodology and Setup}
\label{sec:method}

Our goal is to analyse how speech characteristics are structured in the factorised SSL representations.
We are specifically interested in whether individual characteristics are encoded in specific dimensions.
For the content subspace $\mathbf{C}$, correlation analysis is directly applied between the rank features of $\mathbf{C}$ and the frame-level speech characteristics.
For the speaker subspace $\mathbf{S}$, PCA is applied on flattened matrices to obtain the principal dimensions; correlation analysis is performed between these principal dimensions and the speaker-level averaged speech characteristics.

\subsection{Speaker characteristics}
\label{subsec:speaker_charac}

We consider pitch (F0, Hz), F1--F3 formants (Hz), intensity (dB), spectral rolloff (Hz; the frequency below which 50\% of a frame's energy resides~\cite{scheirer_etal,zcr_spec_roll}), zero-crossing rate (ZCR; an indicator of voicing and noisiness), and gender.
For $\mathbf{S}$ we additionally include timbre characteristics: local jitter and shimmer (\%; indicating pitch and intensity stability, respectively~\cite{jitter_shimmer}) and harmonic-to-noise ratio (HNR, dB; a noise-level measure~\cite{hnr}). HNR and ZCR are also affected by channel properties~\cite{zcr_spec_roll}.
Pitch, formants, intensity, jitter, shimmer, and HNR are calculated with Parselmouth's Praat functions~\cite{praat,parselmouth}; ZCR and spectral rolloff with Librosa~\cite{librosa}.
Characteristics are measured per frame for the $\mathbf{C}$-subspace analysis, and averaged per speaker for the $\mathbf{S}$-subspace analysis.

\subsection{Representations}
\label{subsec:representations}

We use 1024-dimensional SSL features from WavLM layer six, following~\cite{pasad2023,kNN-VC}.
Since each $\mathbf{S}_k$ retains the full 1024-dimensional representation, it complicates correlation analysis.
We therefore flatten and stack all $\mathbf{S}_k$ matrices and apply PCA (using Scikit-learn~\cite{scikit-learn}, with 50 components) to obtain linear, variance-ordered principal dimensions of this flattened $\mathbf{S}$-subspace.
Although non-linear dimensionality reduction approaches could be considered, linear PCA is simple and directly interpretable.

\subsection{Data}
\label{subsec:data}

Stable PCA estimation requires many speakers, so we extract $\mathbf{S}$ from the 1596 speakers in Libri-Light's medium partition~\cite{Libri-Light}.
For correlation and intervention analyses of $\mathbf{S}$ and $\mathbf{C}$, we use the 331 speakers across LibriSpeech's train-clean-100, dev-clean, and test-clean sets~\cite{LibriSpeech}.
$\mathbf{S}$-correlation experiments average characteristics per speaker, so any extraction noise has minimal effect.
$\mathbf{C}$-correlation, however, is frame-level and thus more sensitive to noisy audio. 
So we additionally curate validation and test sets of 100 utterances each (10 speakers $\times$ 10 utterances, 5 male/5 female, from dev-clean and test-clean respectively) to be used as ground-truth measurements, manually verifying their characteristic estimates against spectrograms.

\subsection{Correlation analysis}
\label{subsec:corr_analysis}

We measure the correlation between each dimension and each characteristic using the coefficient of determination $R^2 \in (-\infty, 1]$~\cite{ISL_python} for the continuous characteristics, where 1 indicates a perfect linear relationship, 0 indicates no better than predicting the mean, and a negative score is worse than mean prediction.
Cohen's kappa $\kappa \in [-1, 1]$~\cite{cohen_kappa} is used for the categorical gender label, where 1 indicates perfect agreement, 0 chance-level agreement, and $-1$ is worse than random chance.

\begin{figure*}[t]
\centering
    \begin{subfigure}[b]{0.33\textwidth}
        \centering
        \includegraphics[width=\linewidth]{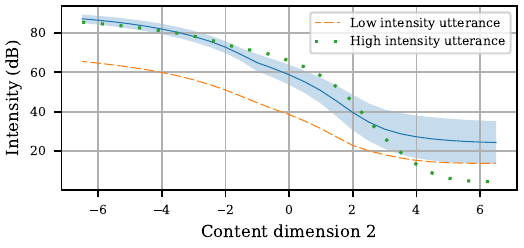}
        \caption{Intensity via content dim. 2.}
        \label{fig:C_intensity_control}
    \end{subfigure}
    \hfill
    \begin{subfigure}[b]{0.33\textwidth}
        \centering
        \includegraphics[width=\linewidth]{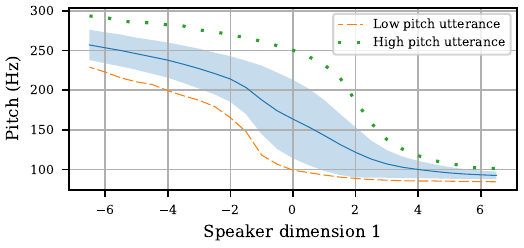}
        \caption{Pitch via speaker dim. 1.}
        \label{fig:S_pitch_control}
    \end{subfigure}
    \hfill
    \begin{subfigure}[b]{0.33\textwidth}
        \centering
        \includegraphics[width=\linewidth]{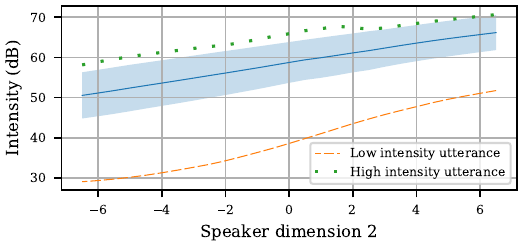}
        \caption{Intensity via speaker dim. 2.}
        \label{fig:S_intensity_control}
    \end{subfigure}
\caption{
The effect on a particular characteristic as its correlated subspace dimension is varied on test data. The blue line shows the average characteristic value as the dimension is changed by a factor of the standard deviation across all test utterances; the shaded area indicates one standard deviation. The dotted-green and dashed-orange lines show changes for individual utterances with, respectively, high and low characteristic values before modification.
}
\label{fig:C_and_S_separate_control_plots}
\end{figure*}

\section{Dimension Analysis}
\label{sec:correlation_analysis}

In this section, we analyse how the rank dimensions of $\mathbf{C}$ (content dimensions) and the principal dimensions of $\mathbf{S}$ (speaker dimensions) correlate with each speech characteristic, considering each subspace independently and leaving joint effects for the next section.

\subsection{Content subspace analysis}
\label{subsec:correlation_content_analysis}

Characteristics are measured frame-wise and correlated with the content dimensions of $\mathbf{C}$.
Fig.~\ref{fig:C_correlations} shows the $R^2$ and $\kappa$ scores for the first 15 (of $r=64$) dimensions on the development set, with the highest-correlating dimension per characteristic outlined in red.

Intensity correlates most strongly overall with dimension 2 ($R^2=0.72$).
The F1, F2, and F3 formants correlate with dimensions 5, 3, and 2 respectively, though only moderately ($R^2$ of 0.136--0.285).
Spectral rolloff and ZCR are both captured by dimension 5, while pitch is captured mainly in the later dimension 12 ($R^2=0.375$).
Gender does not correlate with any dimension, consistent with it being a static characteristic that does not vary temporally within an utterance.

\subsection{Speaker subspace analysis}
\label{subsec:correlation_speaker_analysis}

Characteristics are averaged per speaker and correlated with the speaker dimensions of $\mathbf{S}$ (the PCA directions).
Fig.~\ref{fig:S_correlations} shows $R^2$ and $\kappa$ scores for a subset of the 50 dimensions.
In contrast to $\mathbf{C}$, dimension 1 alone correlates strongly with several characteristics: jitter, pitch, and gender all give $R^2$ and $\kappa$ of more than 0.58.
Formants and intensity are weakly, or not at all, correlated with any dimension.
Dimension 5 correlates somewhat with HNR and shimmer, and dimension 7 with spectral rolloff and ZCR ($R^2=0.32$ and $0.37$), both linked to higher-frequency content.

Comparing $\mathbf{C}$ and $\mathbf{S}$ (Fig.~\ref{fig:correlation}) shows a consistent pattern: characteristics that vary temporally within an utterance, such as intensity and the formants, correlate strongly with $\mathbf{C}$ but weakly or not at all with $\mathbf{S}$, while static characteristics, such as gender, show the reverse.
Spectral rolloff and ZCR, tied to both temporal voicing and static channel properties, correlate with dimensions in both subspaces.
Pitch is a notable exception: it correlates very strongly with a speaker dimension and only weakly with a much later content dimension, consistent with a speaker's average pitch being captured in $\mathbf{S}$ while within-utterance pitch variation is captured in $\mathbf{C}$.

\section{Intervention Analysis}
\label{sec:intervention_analysis}

Turning from correlation measurements to targeted interventions, we present synthesis experiments to measure how the $\mathbf{C}$ and $\mathbf{S}$ subspaces influence the downstream characteristics, both independently and in tandem.
Specifically, we explicitly modify a dimension’s value and track the causal effect on the resulting audio after vocoding the modified SSL feature sequence. For this section, we use the average pitch and intensity as running examples, but the experiments were performed for all speech characteristics.

\subsection{Experimental setup}

We take an inference utterance and determine its $\mathbf{C}_{\text{inf}}$ and $\mathbf{S}_{\text{inf}}$ matrices, as explained in Sec.~\ref{sec:Content_fact}.
We modify one of the content dimension values of $\mathbf{C}_{\text{inf}}$, one of the speaker dimension values of $\mathbf{S}_{\text{inf}}$, or a combination of the two, then resynthesise the audio and measure the resulting speaker characteristics, analysing how they change as the dimensions are modified.

Formally, for modifying dimension $i$ of $\mathbf{C}_{\text{inf}}$, we simply add a constant $\alpha \sigma_i$ to each value in the $i$th column of $\mathbf{C}_{\text{inf}}$, producing $\mathbf{C}_{\text{inf}}^{\text{mod}}$, where $\sigma_i$ is the standard deviation over the dataset of the $i$th dimension in $\mathbf{C}$.
For $\mathbf{S}_{\text{inf}}$, we add a scalar multiple $\alpha$ of the unflattened, unit-length principal direction $\mathbf{V}_{i}$ to $\mathbf{S}_{\text{inf}}$, such that $\mathbf{S}_{\text{inf}}^{\text{mod}} = \mathbf{S}_{\text{inf}} + \alpha \sigma_{i} \mathbf{V}_{i}$, where $\sigma_i$ is the standard deviation for principal dimension $i$.
We then find our modified SSL frame representations, $\mathbf{X}_{\text{inf}}^{\text{mod}}$, by using $\mathbf{C}_{\text{inf}}^{\text{mod}}$ and $\mathbf{S}_{\text{inf}}^{\text{mod}}$ in~\eqref{eq:content_fact}, and synthesise the speech from $\mathbf{X}_{\text{inf}}^{\text{mod}}$ using a pretrained WavLM-to-waveform HiFi-GAN~\cite{HiFi-GAN,kNN-VC}.

\begin{figure*}
\centering
    \begin{subfigure}[b]{0.49\textwidth}
        \centering
        \includegraphics[width=\linewidth]{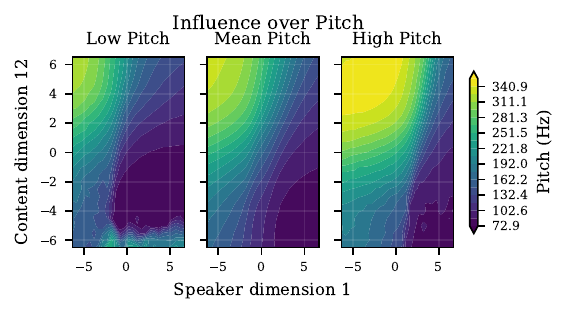}
        \caption{Varying pitch via content dimension 12 and speaker dimension 1.}
        \label{fig:C_and_S_pitch_control}
    \end{subfigure}
    \hfill
    \begin{subfigure}[b]{0.49\textwidth}
        \centering
        \includegraphics[width=\linewidth]{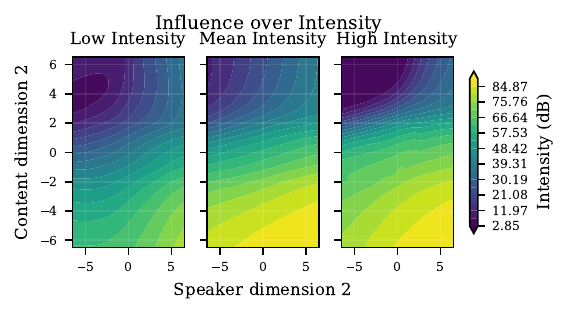}
        \caption{Varying intensity via content and speaker dimensions 2.}
        \label{fig:C_and_S_intensity_control}
    \end{subfigure}
\caption{
The effect on a specific characteristic as its correlated content and speaker dimensions are varied. The middle contours show the average characteristic value as the dimensions are varied across all test utterances. The other two contour plots show the characteristic change for speakers with naturally low (left) and high (right) characteristic values.
}
\label{fig:C_and_S_control_plots}
\end{figure*}

\subsection{Individual subspace analysis}
\label{subsec:synthesis_individual_analysis}

We first perform interventions on $\mathbf{C}$ and $\mathbf{S}$ in isolation to determine how much control a single dimension exerts over a correlated characteristic.
Fig.~\ref{fig:C_and_S_separate_control_plots} shows the resulting average trend (blue line, $\pm 1$ standard deviation shaded) as a content or speaker dimension is varied across all test utterances, together with two individual high- and low-valued utterances (green and orange).
When we qualitatively listen to the manipulated audio, we hear that it also maintains good quality when the modification is within reasonable limits.\footnote{Audio samples: \url{https://kylejvr767.github.io/ICASSP_2027_demo_page/}}
To test manipulation of individual utterances, we also selected individual high- and low-pitched or -intensity test utterances, represented by the green-dotted and orange-dashed lines, respectively.

Pitch correlates only weakly with $\mathbf{C}$ but strongly with speaker dimension 1 (Fig.~\ref{fig:correlation}).
Speaker dimension 1 is able to manipulate the pitch to be between 100 and 250~Hz on average (Fig.~\ref{fig:S_pitch_control}), and content dimension 12 gives comparable control over a 100--300~Hz range despite the weak correlation (not pictured individually, but its contribution alongside the speaker dimension is visible in the joint intervention of Sec.~\ref{subsec:syntehsis_C_and_S_analysis}).
Listening to the $\mathbf{C}$-modified utterances, the pitch changes because the speaking style itself is being changed at a frame level.
Both trends plateau as the dimension is moved excessively far from the data distribution: for the low-pitch example, pushing content dimension 12 below a factor of $-$5 of the standard deviations causes the pitch to suddenly increase again, since significant noise is added by the vocoder once the modified SSL sequence is far out of distribution.
$\mathbf{S}$-space manipulation, by contrast, remains stable across the full $[-6,6]$ range tested, without this vocoder failure.
Speaker dimension 1 also correlates with gender, so large pitch shifts via $\mathbf{S}$ audibly change perceived gender, which we also hear when listening to the synthesised audio; this effect is not observed when shifting pitch through $\mathbf{C}$.

Intensity shows the opposite pattern: it correlates strongly with content dimension 2 but only weakly with any speaker dimension.
Consequently, intensity can be controlled between 20 and 85~dB via $\mathbf{C}$ (Fig.~\ref{fig:C_intensity_control}), with the standard deviation around the average trend much tighter than for pitch. 
This shows greater precision and generalisation of control, but only between 50 and 65~dB via $\mathbf{S}$ (Fig.~\ref{fig:S_intensity_control}), where the trend is linear rather than plateauing.
We again tested all the other characteristics: manipulation is possible for each by altering the corresponding dimension in $\mathbf{C}$, always with the same plateauing trend; pushing to extremes again causes the vocoder to add significant noise, similar to the pitch behaviour above.
Most characteristics can also be modified to some degree using $\mathbf{S}$, but with lesser effect and more noise. This is especially the case for spectral rolloff and ZCR, even though both are correlated to the same extent in both subspaces.
Average pitch was the only characteristic to have a plateauing trend through $\mathbf{S}$, though; all others were linear, and the formants could not be controlled through $\mathbf{S}$ at all.

\subsection{Simultaneous modification of content and speaker spaces}
\label{subsec:syntehsis_C_and_S_analysis}

Finally, we modify $\mathbf{C}$ and $\mathbf{S}$ simultaneously to determine whether they retain their effect when coupled together.
Fig.~\ref{fig:C_and_S_control_plots} shows the results of these simultaneous interventions as heatmap contour plots.

We first consider the effect on pitch (Fig.~\ref{fig:C_and_S_pitch_control}).
On average (middle plot), there is a clear separation between a light and a dark area (lighter colours indicate high pitch values and darker colours low pitch).
The total range that can be achieved is between 70 and 340~Hz, a much wider range than either dimension modified alone (Fig.~\ref{fig:C_and_S_separate_control_plots}).
Neither dimension overpowers the other, but the speaker dimension has the primary influence: keeping the content dimension fixed and varying only the speaker dimension produces a larger range of change than the reverse. 
There are, however, limits, as shown when trying to alter the extreme low- or high-pitched utterances (left and right in the figure): as before, pushing either dimension to extremes can drive the vocoder out of its operating region, seen in the low-pitched utterance when content dimension 12 drops below $\alpha=-0.4$, causing a sudden increase in pitch.

When considering the average intensity (Fig.~\ref{fig:C_and_S_intensity_control}), we again see two extreme areas, but this time the content dimension takes the lead over the speaker dimension, consistent with the correlation analysis (Fig.~\ref{fig:correlation}) and the isolated interventions above (Fig.~\ref{fig:C_and_S_separate_control_plots}).
Nevertheless, using the two subspaces in tandem, we can achieve intensities between 2 and 85~dB, wider than when the $\mathbf{C}$-space is modified in isolation (Fig.~\ref{fig:C_intensity_control}).

We also tested the remaining speech characteristics, and they all show two extreme regions that we can move between with a varying influence balance between the two subspaces, giving relative, wide-range control over the characteristic throughout.

\section{Conclusion}
\label{sec:conclusion}

We investigated how speech characteristics are structured within content and speaker subspaces of SSL representations, factorised via SVD on WavLM features.
Correlation analysis revealed that both subspaces capture distinct characteristics within isolated, individual dimensions, and intervention experiments showed these dimensions can be manipulated, independently and in combination, to control specific characteristics without disrupting lexical content.
This opens up the possibility of voice manipulation through a simple, training-free approach, which we aim to formalise in the~future.

\vspace{5pt}

\noindent \textbf{AI disclosure.} The code for this paper was mainly written by hand. The text was written in expanded form and then shortened using Claude. A complete manual review was then conducted.

\bibliographystyle{IEEEbib}
\bibliography{refs}

\end{document}